\documentstyle[prl,aps,epsf]{revtex}
\draft
\begin{document}
\twocolumn[\hsize\textwidth\columnwidth\hsize\csname @twocolumnfalse\endcsname

\title{Spontaneous Symmetry Breaking For Driven Interacting Particles on 
Triangular Substrates} 
\author{C. Reichhardt and C.J. Olson Reichhardt} 
\address{ 
Center for Nonlinear Studies and Theoretical Division, 
Los Alamos National Laboratory, Los Alamos, New Mexico 87545}

\date{\today}
\maketitle
\begin{abstract}
For collectively interacting repulsive
particles driven on triangular substrates, we show that for certain
directions of drive
a spontaneous symmetry breaking phenomena occurs where the
particles can flow in one of two directions    
that are not aligned with the external drive, giving rise to 
a positive or negative Hall current.
Along these directions, 
the particle flow is highly ordered,  
while in the direction of the drive the flow
is disordered. We also find a number of dynamical 
phase transitions and unusual hysteretic properties that arise
due to the symmetry breaking properties of the flows.  
\end{abstract}
\vspace{-0.1in}
\vspace{-0.3in}

\vskip2pc]
\narrowtext
A number of systems can be represented as collectively 
interacting particles moving over an underlying periodic substrate,
such as models  
of atomic friction \cite{Granato}, 
vortex flow in superconductors with 
periodic arrays of pinning sites \cite{Harada,Reichhardt,Locking}, 
and colloids interacting with periodic optical 
trap arrays \cite{Grier,Grier2,Bechinger}.
Typically these systems exhibit 
a rich variety of 
{\it dynamical} phase transitions under increasing applied drive, 
such as transitions from highly disordered fluctuating flows to
highly ordered one-dimensional (1D) and 2D flows, with strong hysteresis
between the phases 
\cite{Granato,Reichhardt,Locking}.
Other effects such as negative resistance and switching can also occur 
in conjunction with these dynamic transitions \cite{Granato,Reichhardt}. 
Usually in these systems the average drift velocity is finite in the direction
of the applied force but zero in the direction transverse to the force. 
Under certain conditions, however,
the particles do not flow in the
direction of drive, but lock into a symmetry direction of the underlying
lattice. 
For a square substrate lattice,
changing the angle of drive 
with respect to the substrate produces
a series of dynamical 
locking phases forming a Devil's staircase 
structure \cite{Locking}. 
Locking effects have
been demonstrated for vortices moving in superconductors
with periodic pinning arrays in both simulations
\cite{Locking} and experiments \cite{Martin}, 
theoretically for electrons on periodic substrates \cite{Wiersig},
and in experiments \cite{Grier} and theory \cite{Grier3}
for colloids driven over periodic traps,
where it was shown that the locking can be used
to sort particles \cite{Grier4}.
The directional locking effects
are not due to collective interactions 
of the driven particles and can occur for a single driven particle. 

For certain periodic substrate lattice geometries,
there are two symmetry directions of the substrate that the particles
could follow instead of
the direction of drive.
An example is a triangular lattice 
oriented such that the applied drive is between two symmetry 
directions, as illustrated schematically in Fig.~1.
In this work we show that when there are many interacting 
driven particles in such a geometry, 
spontaneous symmetry breaking 
can occur where the
particles preferentially 
flow globally along either of the two symmetry directions rather than in 
the direction of drive. 
Since the substrate is symmetric,
the occurrence of the global flow can be viewed as
a spontaneous symmetry breaking.
This transition is driven by {\it collective interactions} 
between the repulsive 
particles. In the limit of a single particle, symmetry breaking is
not observed. 
When the global flow follows one of the symmetry directions of the substrate,
the flow is ordered with few fluctuations and a uniform spacing between the
particles, reducing their interaction energy.
If the global flow instead follows the direction 
of the drive, the particle
configuration and flow are much more
disordered. 
At $T = 0$ for ordered configurations of the driven particles,
such as at commensurate fillings, the particles
flow in the direction of drive. At incommensurate fillings
or finite temperatures, however,
enough perturbations can occur to allow the system
to order into one of the two easy flow directions. 
For different system 
parameters such as waiting times, ramp rates,
fillings, and initial conditions, 

\begin{figure}
\center{
\epsfxsize=3.5in
\epsfbox{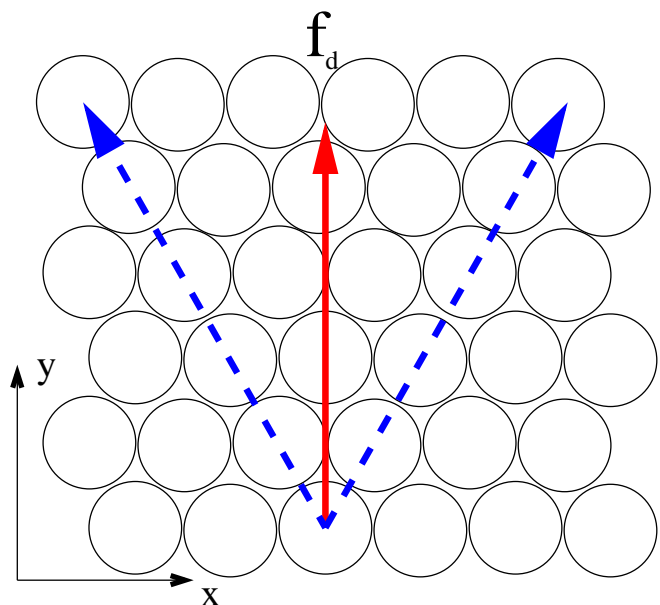}}
\caption{
Schematic of the system with a triangular substrate. The potential
minima are at the center of the circles. Each minima captures one
or more interacting particles. The applied driving force is $f_d {{\bf \hat y}}$
(solid arrow). There are two 
easy flow directions at $\pm30^\circ$ from the drive (dashed arrows).  
}
\end{figure}

\noindent
the symmetry breaking is equally likely to occur in either direction; 
thus, for an ensemble of realizations, the
symmetry is restored in the statistical sense. 
Additionally we find a series of dynamical phase transitions
where the particle flow changes from ordered to disordered. This
can produce unusual hysteretic effects. 
For example, when increasing the drive,
the transverse velocity may lock to the positive direction at low drives,
but if the system 
passes through a disordered flow regime at high drives, then 
when decreasing the drive the flow has an equal probability to 
lock in the negative direction. 
We explicitly show this effect for colloids moving over periodic substrates
and vortices in superconductors with triangular pinning arrays.

We consider a system of $N_{c}$ overdamped repulsively interacting particles 
interacting with an underlying triangular periodic substrate with $N$ minima. 
The equation of motion for a particle $i$ is
\begin{equation}
\frac{d {\bf r}_{i}}{dt} = {\bf f}_{ij} +
{\bf f}_{s} + {\bf f}_{d} + {\bf f}_{T} \ .
\end{equation}
Here the
particle-particle interaction force is
${\bf f}_{ij} = -\sum_{j \neq i}^{N_{c}}\nabla_i V(r_{ij})$,
where we use a Yukawa or screened Coulomb
interaction potential appropriate for colloids, given by
$V(r_{ij}) = (q^2/|{\bf r}_{i} - {\bf r}_{j}|)
\exp(-\kappa|{\bf r}_{i} - {\bf r}_{j}|)$. 
Here  $q$ is the colloid charge, $1/\kappa$ is the 
screening length, and ${\bf r}_{i(j)}$ is the position of
particle $i (j)$.  
We also consider vortices in 
thin film superconductors with periodic pinning.
In this case the vortices interact with a logarithmic potential
$V(r_{ij}) = -\ln(r_{ij})$.  
For the colloids we consider a triangular 
substrate which can capture multiple colloids per minima:
${\bf f}_{s} = \sum_{i=1}^{3}A\sin(2\pi p_{i}/a_{0})
[\cos(\theta_{i}){\hat {\bf x}} - \sin(\theta_{i}){\hat {\bf y}}]$, where 
$p_{i} = x\cos(\theta_{i}) - y\sin(\theta_{i}) + a_{0}/2$, 
$\theta_{1} = \pi/6$, $\theta_{2} = \pi/2$, and $\theta_{3} = 5\pi/6$. 
Such substrates 
can be created with optical trap arrays \cite{Bechinger}
and have been studied in previous simulations \cite{Olson}.
In the case of vortices, the pinning sites are modeled as short range
attractive parabolic pinning sites
in a triangular lattice that capture one vortex each. The
additional vortices sit in the interstitial regions 
between the pinned vortices. 
Length 
is measured in units of the substrate lattice constant 
$a_{0}$ and we use
a screening length for the colloids  of $\kappa = 3/a_{0}$. 
The thermal force ${\bf f}_{T}$ is a randomly fluctuating force 
from random kicks. 
We consider both $T = 0$ and finite $T$. 
The initial particle positions are obtained by simulated annealing
from a high temperature. 
The driving force $f_{d}$ is applied in the $y$ direction
as shown in Fig.~1.
The drive is increased in small increments to a finite value and then
similarly decreased back to zero. 
We measure both the velocity in the direction of 
drive, $V_{y} = \sum_{i}^{N}{\bf v_{i}}\cdot {\bf {\hat y}}$,  
and the velocity transverse to the drive,
$V_{x} = \sum_{i}^{N}{\bf v_{y}}\cdot {\bf {\hat x}}$.  

We first consider the case of colloidal particles interacting with
a triangular substrate. We concentrate on  
a regime where the number of colloids
is greater than the 
number of potential minima, as in 
recent experiments 
on triangular substrates 
\cite{Bechinger}.   
Fig.~1 shows a schematic of 

\begin{figure}
\center{
\epsfxsize=3.5in
\epsfbox{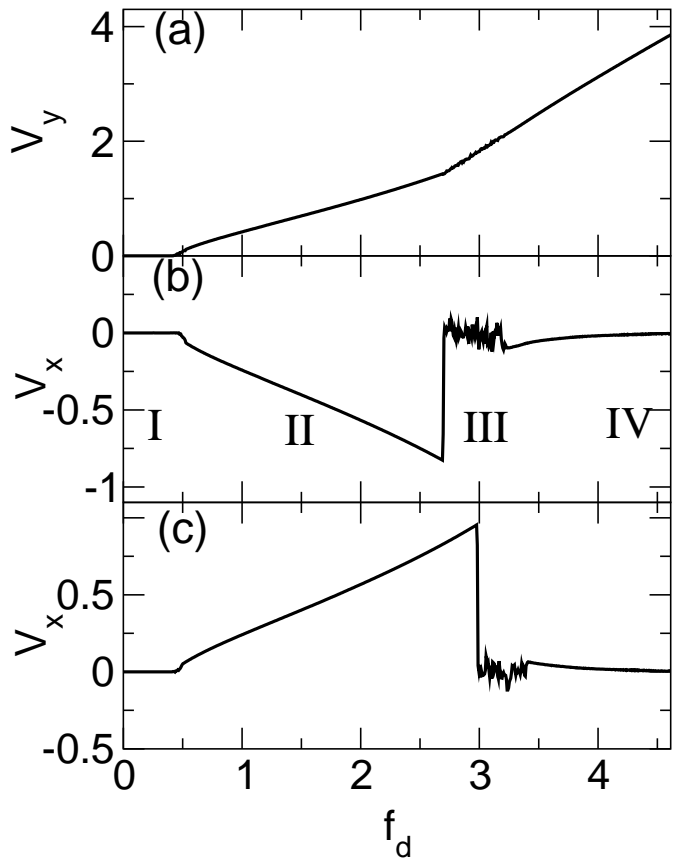}}
\caption{
(a) Velocity in the direction of drive
$V_{y}$ vs applied drive $f_{d}$ for a system of interacting
colloids on a triangular substrate for $N_{c}/N = 1.39$.
(b) Corresponding transverse velocity $V_{x}$ vs $f_{d}$.
Regions I-IV correspond to different particle flow regimes.
(c) The transverse velocity $V_{x}$ vs $f_{d}$ for the same
system as above but for slightly different initial conditions
showing that the transverse velocity reverses from (b). 
}
\end{figure}

\noindent
our system. 
The dashed arrows indicate the
symmetry directions for particle flow. 
In Fig.~2(a) we plot the average particle velocity
in the direction of the drive, $V_{y}$, vs applied force for a system 
with $N_{c}/N = 1.39$. In Fig.~2(b) we show the corresponding 
average velocity in the $x$-direction, $V_{x}$. Here four distinct 
dynamic regions can be 
observed. The first region, I, is the pinned phase where 
particles do not move in either direction. 
Region II is the  
spontaneous symmetry breaking phase where the
global flow of particles follows
either the positive or negative $30^\circ$ direction
as in Fig.~1. 
In this regime $V_{y}$ and $V_{x}$ both increase linearly
with increasing $f_{d}$.
In Fig.~2(b) the particles flow along 
$-30^\circ$, producing a negative $V_{x}$ or Hall current.
As $f_{d}$ increases, the particles start to move
in the direction of the drive 
and there is a transition to region III flow. This
transition appears as a
sharp drop in $V_{x}$ with a corresponding increase in the
slope of $V_{y}$.  
Region III is a disordered flow regime
with zero average flow in the 
$x$-direction. 
At higher drives we find a crossover to
a more ordered flow in Region IV, 
which can be identified by the reduction of fluctuations in $V_{y}$.
In region IV there can be
a slight initial drift in the $x$ direction due to the orientation of
the moving particle lattice
with respect to the triangular substrate,
but as $f_{d}$ further increases, $V_{x}$ goes to zero. 

In Fig.~2(c) we show $V_{x}$ for the same system as in Fig.~2(a,b)
but with slightly different initial particle positions.
Here the curve appears almost the same as in Fig.~2(b), but 
the flow in region II is in the positive $x$-direction. 
We find such
symmetry breaking flow, 
which is equally likely to be in the positive or negative direction, 
for fillings $N_{c}/N > 1.0$.
There is a small 
region below the onset of 
region II where the flow 
shifts transiently between
the positive and negative directions before it 
locks to one of the directions.
We have done simulations for larger systems and find the same type
of velocity-force curves;  however, the transitory time for
the flow to organize into region II increases with system size.  
At $N_{c}/N = 2$ and $3$ and $T = 0$, region II does not occur
because the particle configurations are completely ordered
at these commensurate 
fillings and there are no perturbations to knock the system
into one of the symmetry directions of the lattice.  At finite temperatures,
however, 
region II reappears for these fillings.
At incommensurate fillings there is a symmetry breaking due to the
positional disorder of the particles. 
For fillings $1.0 < N/N_{c} < 2.0$ 
the  velocity force curves look very similar
to the curves shown in Fig.~2. For 
$N/N_{c} > 2.0$, region II can still occur; however, additional dynamical
phases arise which appear as features in the velocity force
curves and the noise fluctuations.  
We have also studied the temperature dependence of the 
phases in Fig.~2. We find that there is a sharp transition temperature 
at which region II ends. This temperature is about $0.7 T_d$, where
$T_d$ is the onset temperature for particle diffusion at 
$f_{d} = 0$. In the inset of Fig.~4, we plot $V_{x}$ for
fixed $f_{d}=1.0$ as a function of temperature showing the two branches of the 
velocity flow up to $T_{c}$ where region II is lost. The sharp loss
of region II as a function of $T$ is consistent with the
sharp transition from region II to the disordered region III.     

In Fig.~3 we plot the particle positions and trajectories
for the different phases. Fig.~3(a) shows the pinned
phase for the system in Fig.~2, where a portion of 
the sites capture two or more colloids. In Fig.~3(b) 
region II is shown for Fig.~2(b) at $f_{d} = 1.0$, with the
particles moving in 1D paths along $-30^\circ$.
In Fig.~3(c), the case for Fig.~2(c) is shown for the same $f_{d}$ with the
flow along $+30^\circ$. Fig.~3(d) illustrates region III flow
from Fig.~2(c) at $f_{d} = 3.2$,
where the particle positions and trajectories are
disordered.
Region IV flow at
higher $f_{d}$ 
is much more ordered and the
particles move mostly 
straight along the $y$ direction; however, there are still some 
dislocations which cause the trajectories to be partially disordered.  
 
The region II flow can be regarded as an example of 
spontaneous symmetry breaking. 
There is no asymmetry 
across the $x$-axis and
in any given realization the
system is equally likely to lock to either
$+30^\circ$ or $-30^\circ$. 

\begin{figure}
\center{
\epsfxsize=3.5in
\epsfbox{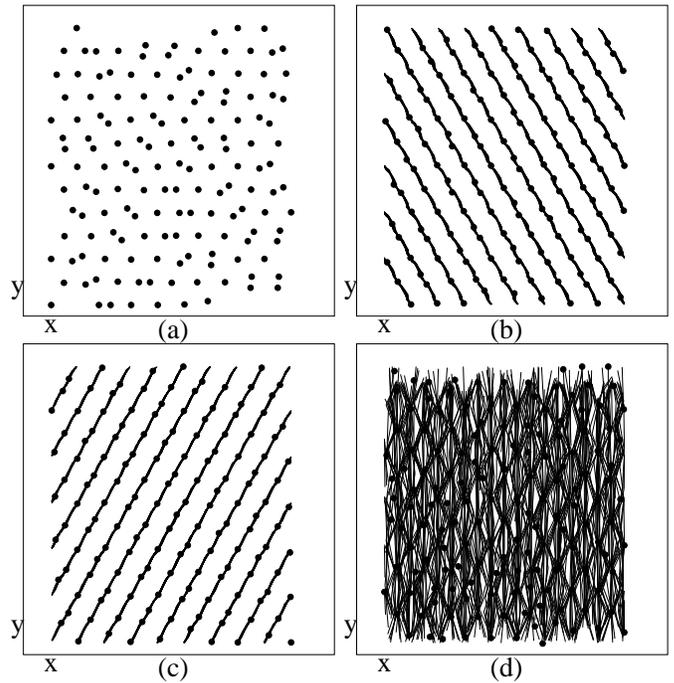}}
\caption{
Colloid positions (black dots) and trajectories (black lines) 
for the system in Fig.~2. (a) Pinned region I, $f_{d} = 0.$
(b) $-30^\circ$ symmetry breaking flow at $f_{d} = 1.$
(c) $+30^\circ$ symmetry breaking flow at $f_{d} = 1.$
(d) Disordered flow in region III, $f_{d} = 3.2$.
}
\end{figure}

\noindent
Spontaneous symmetry breaking
in general is not limited to equilibrium systems.    
In condensed matter, the most familiar example is
that of a ferromagnet
in zero external field where there are two 
energy minima.  A small perturbation from temperature or dynamics
biases the system in one direction or another.  
In our system, when the flow 
follows one of the symmetry directions,
the flow is more ordered and the particle spacing is more uniform.

We have also considered a single particle for the
system in Fig.~1 as well as other fillings with  $N_{c}/N < 1.0$.
In these cases the global flow locking does not occur. 
This is due to the lack of interactions at these low fillings. At depinning,
each particle can move in either the positive or negative direction at each
potential maximum. If there are no neighboring particles blocking 
one of the routes, the particle flows in a random zig-zag pattern.  

Some of the features in the velocity force curves in Fig.~2 can be understood
by force balance arguments. For the commensurate
case $N_{c} = N$, the colloid configurations are 
completely symmetric and the interaction forces are zero, so 
depinning occurs at $f_{d} = f_{s}^{max}$, 
where 
$f_{s}^{max}=3.0$ 
is the maximum force force from the substrate.
For 
higher $N_{c}$, some substrate minima capture more than one colloid. 
In these dimerised phases, a 
colloid feels an 
additional force from the other colloid in the
same minima. This will reduce the force
needed to depin 
the colloids by the
interaction force $f_{ij}(a_0/2)$. Using our parameters this would
give a depinning force of $0.8$ which 

\begin{figure}
\center{
\epsfxsize=3.5in
\epsfbox{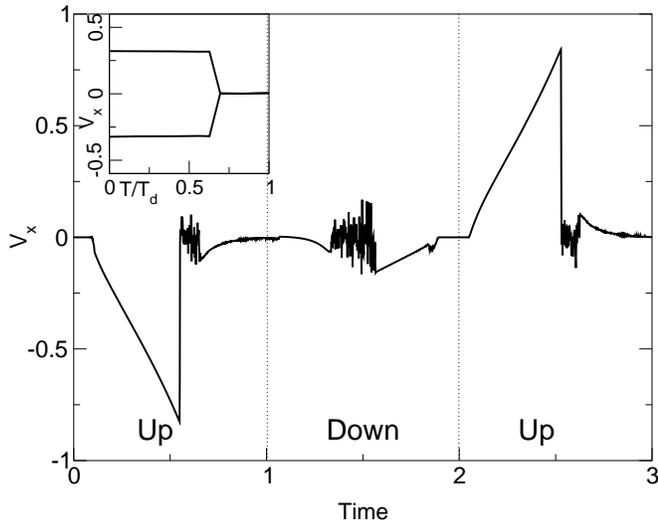}}
\caption{
The transverse velocity $V_{x}$ for the ramp up phase 
of $0.0 < f_{d} \le 4.8$ for time $0<t/t_r<1$; down from $f_{d} = 4.8$ to $0$, 
$1<t/t_r<2$; and
up again to $f_{d} = 4.8$, $2<t/t_r<3$. 
Here $t_r=2.24 \times 10^6$ molecular dynamics
time steps.
Inset: transverse velocity for fixed 
$f_{d}=1.0$ vs $T/T_d$. 
}
\end{figure}

\noindent
is  somewhat higher than 
the measured value
due to the collective interactions of many dimers.     
Fig.~3(a) shows that the dimerised states preferably orient either
in the positive or negative $30^\circ$ direction, so that is the
direction in which they depin. The transition to region IV flow occurs for
$f_{d} > f_{s}^{max}$ where the colloids can move freely over the 
substrate potential maxima.

We next consider the hysteretic properties of the phases by performing 
simulations where the external drive is cycled up, down, and up again. 
In Fig.~4 we show the results for the same system as in Fig.~2 
for three drive cycles.  
For $0<t/t_r<1$ where 
$t_r=2.24 \times 10^6$ molecular dynamics 
time steps, $f_d$ is increased from
zero to 
$f_{d} = 4.8$. 
On the ramp down, beginning at $t/t_r=1$, 
there is clear hysteresis in the III-II transition with
region III persisting longer on the downward sweep. Hysteresis across 
the II-III transition is consistent with the first 
order-like sharp jump in $V_x$ across the II-III transition. 
On the ramp down, region II is again in the negative
direction. The slope is also smaller as
there are more commensurate pinned rows. We note that because
the system goes through region III the flow is disordered so that the
pinned phase near the end of the ramp down phase is not necessarily 
the same as the initial pinned phase. On the next ramp up,
beginning at $t/t_r=2$, $V_{x}$ 
in region  II reverses and is in the positive direction. For 
continued cycling, even at $T = 0$, the direction of the region II flow 
falls randomly in either direction. This is due to the 
dynamical disordering introduced by the region III flow, 
which destroys the memory of the
previous pinned phase.   
We observe similar hysteresis effects for other fillings in the range
$1 < N_c/N < 4$.

In addition to the colloidal system, essentially the same situation 
occurs for vortices driven in superconductors with periodic pinning arrays.
In this case we considered each pinning site to capture one vortex. At
fields with more than two vortices per pinning site, the
same dimer states occur for vortices in the interstitial regions between
the pinned vortices and the same phases arise.  

To summarize, we have  found a novel spontaneous symmetry breaking
phenomena for collectively interacting driven particles on triangular 
substrates. The global particle flow occurs along one of two  
symmetry directions and not in the direction
of the applied drive.
This appears as a positive and a negative finite Hall velocity. 
The flow along the symmetry directions is much more ordered than the flow
in the direction of the drive.
We also find a series of dynamical phase transitions between ordered
and disordered flows which produce jumps in the
longitudinal and transverse velocity force
curves. 
The symmetry breaking combined with the
flow transverse to the drive 
produces interesting hysteresis phenomena in 
the transverse velocity curves where 
the particles may lock to different directions upon increasing and decreasing
the drive, as long as
the particles pass through a disordered flow regime.
The results in this work can be realized for colloids driven over
optical trap arrays and vortices in superconductors with periodic pinning
arrays. Similar effects may be possible for atomic friction systems.

This work was supported by the U.S. DoE under Contract No.
W-7405-ENG-36.


\vspace{-0.1in}
\begin{references}
\vspace{-0.4in}

\bibitem{Granato}
B.N.J.~Persson, Phys.~Rev.~Lett.~{\bf 71}, 1212 (1993); 
O.M.~Braun {\it et al.}, 
{\it ibid}.
{\bf 78}, 1295 (1997);
E.~Granato and S.C.~Ying, 
{\it ibid}.
{\bf 85}, 5368 (2000). 


\bibitem{Harada}
M.~Baert {\it et al.}, Phys.~Rev.~Lett.~{\bf 74}, 3269 (1995);
K.~Harada {\it et al.}, Science {\bf 274}, 1167 (1996);
J.I.~Mart{\' \i}n {\it et al.}, Phys.~Rev.~Lett.~{\bf 83}, 1022 (1999).

\bibitem{Reichhardt}
C. Reichhardt, C.J. Olson, and F.~Nori,
Phys.~Rev.~B {\bf 58}, 6534 (1998);
B.Y.~Zhu {\it et al}, {\it ibid.}
{\bf 64}, 012504 (2001).

\bibitem{Locking}
C. Reichhardt and F. Nori,
Phys.~Rev.~Lett.~{\bf 82}, 414 (1999). 

\bibitem{Grier}
P.T.~Korda, M.B.~Taylor, and D.G.~Grier,
Phys.~Rev. Lett.~{\bf 89}, 128301 (2002).

\bibitem{Grier2} 
P.T.~Korda, G.C.~Spalding, and D.G.~Grier,
Phys.~Rev.~B {\bf 66}, 024504 (2002).

\bibitem{Bechinger}
M.~Brunner and C.~Bechinger,
Phys.~Rev.~Lett.~{\bf 88}, 248302 (2002).

\bibitem{Martin} 
J.E.~Villegas {\it et al.}, cond-mat/0307432, to appear in Phys.~Rev.~B;
A.V.~Silhanek {\it et al.}, cond-mat/0302173, to appear in Phys.~Rev.~B. 

\bibitem{Wiersig}
J. Wiersig and K.-H. Ahn, Phys. Rev. Lett. {\bf 87}, 026803 (2001).

\bibitem{Grier3}
A. Gopinathan and D.G. Grier, cond-mat/0311117. 

\bibitem{Grier4}
K.~Ladavac, K.~Kasza, and D.G.~Grier, cond-mat/0310396. 

\bibitem{Olson}
C.~Reichhardt and C.J.~Olson, Phys.~Rev.~Lett.~{\bf 88}, 248301 (2002).

\end{references}
\end{document}